\documentclass[aps,
 prl,
 reprint,
superscriptaddress,
nofootinbib,
nobibnotes,
amsmath,amssymb,
nolongbibliography,
noeprint,
floatfix
]{revtex4-2}
\usepackage{caption}
\usepackage{soul}
\usepackage{ragged2e} 
\usepackage{graphicx}
\usepackage{mathrsfs} 
\usepackage{amsthm}
\usepackage{physics}
\usepackage[pdftex,colorlinks=true,linkcolor=blue,citecolor=blue,urlcolor=blue]{hyperref}

\usepackage{tikz}
\makeatletter
\renewcommand{\@makecaption}[2]{%
  \vskip\abovecaptionskip
  \begin{minipage}{\linewidth} 
  \small
  \parbox{\linewidth}{\justify #1. #2} 
  \end{minipage}
  \vskip\belowcaptionskip
}
\makeatother
\usepackage{placeins}
\usepackage{wrapfig}
\usepackage{hyperref}
\newcommand{\sect}[1]{\par\textit{#1} --- }

\newcommand{\E}{\mathscr{E}}
\newcommand{\AM}{\mathscr{L}}

\newcommand{\K}{\mathscr{K}}

\newcommand{\consts}{\mathcal{C}}

\begin{document}

\preprint{APS/123-QED}

\newcommand{\LL}[1]{{\textcolor{blue}{#1}}}
\title{Devoured by a Hairy Gargantua: Probing Massive Scalar Charges with \texorpdfstring{\\}{ } Non-minimal Curvature Coupling with Extreme-Mass-Ratio Inspirals}
\author{Leif Lui}
\affiliation{Beijing Institute of Mathematical Sciences and Applications, Beijing 101408, China}

\author{Adrian Ka-Wai Chung}
\affiliation{DAMTP, Centre for Mathematical Sciences, University of Cambridge, Wilberforce Road, Cambridge CB3 0WA, United Kingdom}

\author{Alejandro Torres-Orjuela}
\email[Corresponding author: ]{atorreso@bimsa.cn}
\affiliation{Beijing Institute of Mathematical Sciences and Applications, Beijing 101408, China}

\date{\today}

\begin{abstract}
Ultralight scalar fields nonminimally coupled to curvature can endow rotating black holes with stationary hair and alter nearby orbits. Using recently constructed hairy black-hole spacetimes, we model the transition to plunge of an extreme-mass-ratio inspiral. For scalar masses $\mu M=0.01$ and $0.1$ and dimensionless curvature coupling $\zeta=10^{-3},10^{-4}$, we find $\mathcal{O}(10)$ rad of gravitational-wave dephasing relative to general relativity over $\mathcal{O}(10^3)$ orbits. A multimode Fisher forecast at $\mu M=0.2$, where neighbouring hair solutions permit a numerical $\mu$-derivative, suggests that $\zeta$ and $\mu$ could be measured to precisions of $\simeq0.2\%$ and $0.6$--$1.1\%$, respectively, for a source with signal-to-noise ratio of about 80 in LISA.
\end{abstract}

\maketitle

\sect{Introduction}Introducing an ultralight scalar degree of freedom is a simple and well-motivated way to extend General Relativity (GR) and the Standard Model~\cite{Weinberg:1977ma, Damour:1994zq, Arvanitaki:2009fg, Barack:2018yly, Berti:2015itd, Yunes:2024lzm}. Depending on its mass and interactions, such a field may constitute dark matter or dark energy, act as an inflaton~\cite{Martin:2013tda} or an axion-like particle~\cite{Hook:2018dlk}, and arise in low-energy effective descriptions of grand unified theories or candidate theories of quantum gravity~\cite{Cano:2021rey, Alexander:2009tp, Yunes:2009hc}. Of particular relevance to strong-field gravity, nonminimal couplings between scalar fields and spacetime curvature can evade the assumptions underlying standard no-hair theorems, allowing black holes (BHs) to support nontrivial scalar configurations. These configurations constitute scalar hair and may be characterized by effective scalar charge. Through their backreaction, they modify the exterior geometry and hence the motion of nearby bodies.

Massive scalar fields can also grow macroscopic bosonic clouds around spinning BHs via superradiant amplification~\cite{Starobinskii:1973vzb, Brito:2015oca, Baryakhtar:2020gao}, but this requires the instability to operate for long enough. The curvature-coupled fields considered here need no such growth, as the invariant sources the scalar directly~\cite{Chung2026}, extending the phenomenology of BH scalar hair beyond the superradiant regime.

Extreme-mass-ratio inspirals (EMRIs) provide powerful laboratories for probing such curvature-coupled scalar fields and constraining their properties~\cite{Yunes2012, Maselli:2021men, Barsanti:2021ydd, Barsanti:2022vvl, Fan2026, Speri2026, Barsanti2026, Keijzer_2026}. Since an EMRI can remain in the observational band of a space-based gravitational-wave (GW) detector (i.e. LISA~\cite{LISA_2022, LISA_2022b, LISA_2022c, LISA_2023, LISA_2024}, TianQin~\cite{TianQin_2015, TianQin_2021, TianQin_2024, Torres-Orjuela_2024}, and Taiji~\cite{taiji_2015,taiji_2021}) for months to years, even small departures from vacuum GR can accumulate into measurable changes in its GW signal. Scalar fields can affect an EMRI through conservative modifications of the central BH geometry, additional channels of scalar radiation, or interactions with an ambient scalar environment, including dynamical friction. Here, we focus primarily on the conservative deformation of the primary BH, which modifies the sequence of strong-field orbits, the properties of the innermost stable circular orbit (ISCO), and the subsequent transition-to-plunge phase. The large number of accumulated orbital cycles makes EMRIs particularly sensitive to such conservative deformations~\cite{Gair2013, Maselli2022, Zhang2023}, while the transition to plunge provides a complementary endpoint signature from the deepest strong-field region~\cite{Ori2000, Compere2020, Becker2025, Becker2026}.

In this \textit{Letter}, we show that massive scalar hair induced by nonminimal curvature couplings leaves characteristic imprints on the late inspiral and transition-to-plunge dynamics of an EMRI. We consider circular, equatorial trajectories and their associated GW signals. This idealized configuration is astrophysically relevant to an important subset of EMRIs formed in active galactic nucleus disks, where gas interactions can circularize the orbit and align it with the disk, while GW radiation further suppresses the eccentricity during the late inspiral~\cite{Kocsis_2011, Pan2021, Lyu2026}. Specifically, for a representative system with primary mass $M=10^6M_{\odot}$, spin $a=0.8M$, mass ratio $\eta=10^{-4}$, and coupling strength $\zeta=10^{-3},10^{-4}$, we find that the scalar-induced deformation produces an accumulated orbital dephasing of $\delta\varphi \simeq +21.3$ rad for dynamical Chern-Simons (dCS) and $-45.2$ rad for scalar Gauss-Bonnet (sGB) at $\zeta = 10^{-3}$, over the final $\sim1.7\times10^{3}$ orbits of evolution for a scalar-field mass $\mu M = 0.01$ ($\mu\hbar = 1.3\times10^{-18}$ eV). Example trajectories of equatorial transition-to-plunge mergers can be found in the Supplementary Material.

\sect{Metric Perturbations and Transition-to-Plunge Trajectory Modified by Massive Scalar Charges}Following Ref.~\cite{Chung2026}, we consider two massive (pseudo)scalar fields that couple linearly to a curvature invariant. We work in natural units, $c=G=\hbar=1$. The dynamics are governed by the extension of the Einstein--Hilbert Lagrangian
\begin{equation}\label{eq:lagrangian}
\begin{split}
16\pi\mathscr{L}={}&R
+\ell_1{}^2\vartheta_1\mathscr{G}
+\ell_2{}^2\vartheta_2\mathscr{P}
-\frac{1}{2}\nabla_\nu\vartheta_1\nabla^\nu\vartheta_1
\\
&-\frac{1}{2}\nabla_\nu\vartheta_2\nabla^\nu\vartheta_2
-\frac{1}{2}\mu_1{}^2\vartheta_1{}^2-\frac{1}{2}\mu_2{}^2\vartheta_2{}^2,
\end{split}
\end{equation}
where $\mathcal{R}$ is the Ricci scalar, $\vartheta_{1,2}$ are scalar and pseudoscalar fields with masses $\mu_{1,2}$ and coupling length scales $\ell_{1,2}$. 
$\mathscr{G}=\mathcal{R}^2-4\mathcal{R}_{\mu\nu}\mathcal{R}^{\mu\nu}+\mathcal{R}_{\mu\nu\rho\sigma}\mathcal{R}^{\mu\nu\rho\sigma}$ is the GB invariant and $\mathscr{P}=\tilde{\mathcal{R}}^{\mu\nu\rho\sigma} \mathcal{R}_{\mu\nu\rho\sigma}$ is the Pontryagin density, where $\mathcal{R}_{\mu\nu}$ and $\mathcal{R}_{\mu\nu\rho\sigma}$ are the Ricci and the Riemann tensors respectively, and $\tilde{R}_{\mu\nu\rho\sigma}$ is the dual Riemann tensor. 
The limits $(\ell_1,\ell_2)=(\ell,0)$ and $(0,\ell)$, respectively, recover the sGB and dCS couplings. 
Such curvature couplings are well-motivated by the low-energy limit of string theory~\cite{Cano:2021rey}, loop quantum gravity~\cite{Alexander:2009tp}, baryogenesis, and inflationary axidilaton models~\cite{Kallosh:2022vha}. 
To preserve the action's overall parity invariance, the scalar field must match the parity of the invariant to which it couples. Therefore, sGB gravity requires a parity-even scalar (coupling to $\mathscr{G}$), while dCS gravity requires a parity-odd pseudoscalar (coupling to $\mathscr{P}$).

A stationary BH sources the corresponding scalar charges through non-minimal coupling. The resulting scalar profile back-reacts on the background BH geometry, leading to modifications from the Kerr spacetime that depend on the coupling length scales~\cite{Chung2026}. For an EMRI with a primary rotating BH dressed in a massive scalar hair, the metric can be expressed as a double power series, in terms of the mass ratio $\eta=m_s/M$ and coupling parameter $\zeta=\left(\ell/M\right)^4$
\begin{equation}\label{eq:metric_PS}
g_{\alpha\beta}=\sum_{n,m}\eta^n\zeta^m g_{\alpha\beta}^{(n,m)}, 
\end{equation}
where $M$ and $m_s$ are the masses of the primary and secondary BHs, respectively. 
In Eq.~\eqref{eq:metric_PS}, $g_{\alpha\beta}^{(0,0)}$ is the Kerr metric and $g_{\alpha\beta}^{(0,1)}$ is the stationary, axisymmetric deformation sourced by the primary's scalar hair. The explicit expressions of $g_{\alpha\beta}^{(0,1)}$ and all numerical validations are given in the Supplementary Material. The deformation $g_{\alpha\beta}^{(0,1)}$ leads to a displacement of the ISCO, $\delta r_{\mathrm{ISCO}}$. 

\begin{figure}[t]
\includegraphics[width=\columnwidth]{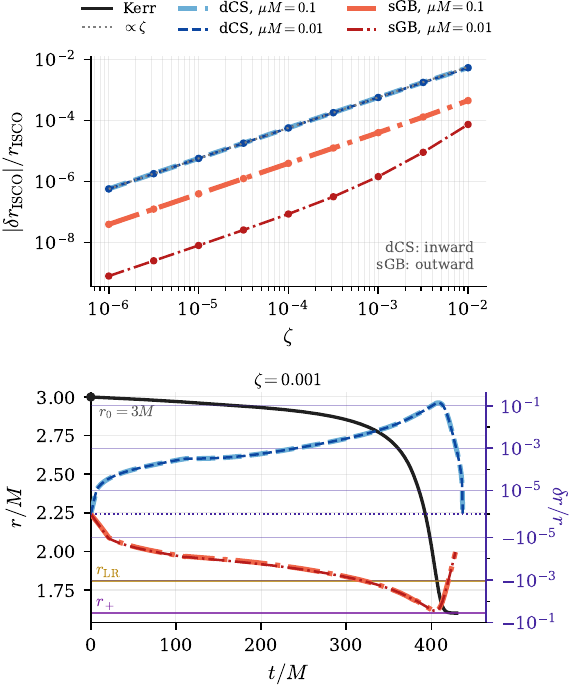}
\caption{\label{fig:isco}\emph{Top}: Magnitude of the fractional shift to the ISCO against the dimensionless coupling parameter of the non-minimal couplings of the massive scalar charges, $\zeta$, at $a=0.8M$, for $\mu M=0.01$ (thin) and $0.1$ (thick). \emph{Bottom}: The baseline radial evolution $r(t)$ for the Kerr spacetime (left axis) alongside the relative radial deviation $\delta r/r$ of the modified theories from this baseline (right axis) at $\zeta=10^{-3}$, starting from a common initial orbital separation $r_0=3M$. The line styles used are the same as in the top figure.}
\end{figure}

While ground-based detectors like LIGO, Virgo, and KAGRA have constrained both massless and massive scalar-tensor theories in the stellar-mass regime~\cite{Perkin2021, LIGO_TGR2021, Xie2025}, EMRIs probe a fundamentally distinct, ultralight scalar regime ($\mu \sim 10^{-18}$ eV) that remains unconstrained by current observations. We adopt $\zeta = 10^{-3}$ and $10^{-4}$ as phenomenological benchmarks that balance perturbative validity with observational detectability. Similarly, we choose $\mu M = 0.01$ and $0.1$ to ensure the scalar field's Compton wavelength remains dynamically relevant to the strong-field binary. The resulting fractional ISCO shift is shown in FIG.~\ref{fig:isco}. For $a=0.8M$, $\zeta=10^{-3}$ and scalar mass $\mu M=0.01$ we find $\delta r_{\rm ISCO}/r_{\rm ISCO}=-5.63\times10^{-4}$ for dCS coupling and $+1.4\times10^{-6}$ for sGB coupling. Because the Boyer-Lindquist radius is coordinate-dependent, we quote below the shift of the observable ISCO frequency, $\delta\Omega_{\rm ISCO}/\Omega_{\rm ISCO}=-2.33\times10^{-4}$ and $+6.11\times10^{-4}$ for dCS coupling and sGB coupling, respectively. 
Interestingly, with dCS coupling the ISCO is pushed inwards, which causes the secondary to linger and complete more cycles than in Kerr, while with sGB coupling the ISCO is pushed outwards and the plunge is accelerated (c.f. FIG.~\ref{fig:isco}). 
The sign is therefore a discriminant between the two couplings, not merely a magnitude. Raising the scalar mass to $\mu M=0.1$ changes this frequency shift by $8.7\%$ with dCS coupling and by $26.9\%$ with sGB coupling, so $\mu$ is a weaker handle in the parity-odd theory than in the parity-even one. Both cases have $\mu M\ll1$, and the dependence should sharpen as $\mu M\to1$.

Expanding the $4$-velocity and connection in the same double series, each order obeys an inhomogeneous geodesic equation on the Kerr background,
\begin{equation}\label{eq:forced}
u_{(0,0)}^{\nu}\nabla^{(0)}_{\nu}u^{\alpha}_{(n,m)}=-\Gamma^{\alpha\,(n,m)}_{\rho\sigma}u^{\rho}_{(0,0)}u^{\sigma}_{(0,0)}\equiv f^{\alpha}_{(n,m)},
\end{equation}
so that the constants of motion $\consts\in\{\E,\AM,\K\}$ drift as
\begin{equation}\label{eq:hier}
\frac{\dd\consts}{\dd\lambda}=\eta\left.\frac{\dd\consts}{\dd\lambda}\right\vert_{(1,0)}+\zeta\left.\frac{\dd\consts}{\dd\lambda}\right\vert_{(0,1)}+\mathcal{O}(\eta^2,\eta\zeta,\zeta^2),
\end{equation}
where $\lambda$ is the Mino-Carter time~\cite{Mino2003, Fujita_2009}, $\dd\E/\dd\lambda=-g^{(0,0)}_{t\alpha}f^{\alpha}$, $\dd\AM/\dd\lambda=g^{(0,0)}_{\phi\alpha}f^{\alpha}$, and $\dd\K/\dd\lambda=0$ for equatorial orbits. 
The $(1,0)$ term encapsulates the effects of adiabatic first-order self-forces (0th post-adiabatic order, 1st self-force order; 0PA-1SF) gravitational self-forces in Kerr spacetimes without scalar charges, and the $(0,1)$ term encapsulates the effects due to the modifications of BH geometry due to the massive scalar charge.

\begin{figure}[t]
\includegraphics[width=\columnwidth]{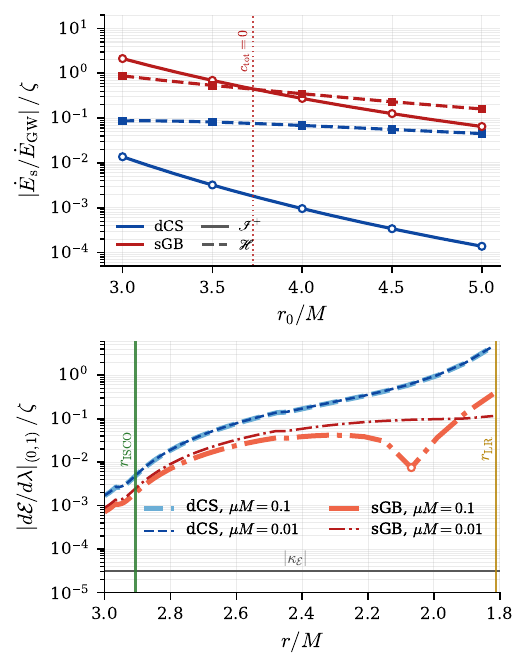}
\caption{\label{fig:radiation}\emph{Top}: the computed $\mathcal{O}(\eta\zeta)$ scalar radiation from the primary's hair at $\mu M=0.01$, per unit $\zeta$ and normalized to the Kerr flux at the same $r_{0}$. Solid curves carry flux out to future null infinity $\mathscr{I}^{+}$, while dashed curves carry it down the horizon $\mathscr{H}$ and are negative. The two channels cancel at $r_{0}\simeq3.73M$ for sGB (dotted line). \emph{Bottom}: the hair-induced forcing per unit $\zeta$ along the worldline, compared against the standard Kerr dissipative radiation reaction $\kappa_{\mathcal{E}}$ at the ISCO ($\eta=10^{-4}$). The line styles used are the same as in the top FIG.~\ref{fig:isco}. The open circle on the sGB $\mu M=0.1$ curve marks the zero-crossing in the forcing term, which manifests as a downward spike on the absolute-value logarithmic scale.}
\end{figure}

For circular equatorial orbits, $f^{t}_{(0,1)}=f^{\phi}_{(0,1)}=0$ identically, as stationary, axisymmetric deformations retain both Killing vectors and therefore cannot do secular work on $\E$ or $\AM$~\cite{Ryan1995, Collins2004, Hinderer2008, Brink2010, Vigeland2010, Vigeland2011, Gair2011}. The scalar-charge-induced force is conservative at $\mathcal{O}(\zeta)$, with dissipation entering only at $\mathcal{O}(\eta\zeta)$ through the scalar channel computed below, and is dynamically suppressed throughout the adiabatic inspiral. During the transit-to-plunge phase, the radial motion is no longer adiabatic, and the scalar-charge-induced force term grows by orders of magnitude as the secondary plunges toward the horizon of the primary (c.f. FIG.~\ref{fig:radiation}). The scalar-charge signature is thus concentrated in the final cycles, precisely where astrophysical environmental contamination is weakest.

A stationary configuration cannot radiate, so the primary's charge can only modify radiation sourced by the secondary, which is itself $\mathcal{O}(\eta)$. Every dissipative channel it opens therefore enters at $\mathcal{O}(\eta\zeta)$, and at $\mathcal{O}(\zeta)$ the hair acts purely conservatively. 

The secondary's metric perturbation stirs the stationary hair, and the disturbed hair radiates. We solve the resulting sourced scalar equation by Green's function, following the treatment of an EMRI inside a scalar cloud in Ref.~\cite{LiWeller2025} and adding the one term our non-minimal coupling demands, which is what makes the source gauge covariant. The top panel of FIG.~\ref{fig:radiation} shows the same competition found there between a positive flux to infinity and a negative, superradiant horizon flux. 

The metric deformation changes the radiative rates by a fractional amount proportional to the coupling, $ c\zeta=\delta\dot{\consts}/\dot{\consts},$ which defines a dimensionless coefficient $c$ for each contribution separately. For clarity, at $\zeta=10^{-3}$, $c=1$ corresponds to a $0.1\%$ change in the flux. Three contributions enter
at $\mathcal{O}(\eta\zeta)$. One, the chain-rule term, $c\simeq2.6$, follows from evaluating the Kerr flux at the orbit displaced by the hair, and we compute it exactly. Two, the modified Teukolsky correction to wave generation, $c\simeq1.3$ (dCS) and $0.9$ (sGB), is the only piece we truncate, and we bound it in the Supplementary Material. Third, the scalar channel shown in FIG.~\ref{fig:radiation}: $c_{\rm s}\le1.26$ for sGB and $\le0.08$ for dCS. The total change in the radiative rates imparts a $3\%$ correction to the chain-rule term in the parity-odd case but roughly half of it in the parity-even one. We compute it and retain it in full, to capture the complete $\mathcal{O}(\eta\zeta)$ scalar dissipation. The secondary's own scalar charge supplies a complementary, dissipative channel of $-1$ pN order, strongest at wide separations where our system is weakest~\cite{Barsanti2026, Speri2026}. Since this probes the secondary rather than the primary, we do not compute it here. The calculations of the scalar radiation and its consistency check with Refs.~\cite{LiWeller2025, Barsanti2026} are provided in the Supplementary Materials. 

Near the ISCO the radial potential $V_{R}(r)$ satisfies $V_{R}=\partial_r V_{R}=\partial_r^2 V_{R}=0$, so expanding to cubic order in $x=r-r_{\rm ISCO}$ while letting the constants drift linearly in Mino time, $\lambda$, reduces the radial equation to~\cite{Ori2000, Becker2025, Becker2026}
\begin{equation}\label{eq:transition}
\frac{\dd^2x}{\dd\lambda^2}=-Ax^2+B\,(\lambda-\lambda_{\rm ISCO}),
\end{equation}
with $A=-\partial_r^3V_{R}/4|_{r=r_{\mathrm{ISCO}}}$ and $B$ fixed by the drift of the constants of motion at the ISCO, $\kappa_{\consts}=\dd\consts/\dd\lambda|_{r=r_{\mathrm{ISCO}}}$, from Eq.~\eqref{eq:hier}. 
We recover $A=M$ for every Kerr spin, and a transition width scaling as $\eta^{2/5}$, in agreement with Refs.~\cite{Ori2000, Becker2025, Becker2026}. 
The $\mathcal{O}(\eta\zeta)$ cross-coupling enters only through $B$, and splits into two parts. The first is the shift of the Kerr flux evaluated at the displaced orbit, $\delta \dot{\consts}_{\rm GW}=(\partial \dot{\consts}_{\rm GW}/\partial\consts)\,\delta\consts$, which we retain exactly~\cite{Kejriwal2024, Teukolsky1973, Drasco_2006}. The second is the explicit modification of the wave generation itself. Since the exact modified Teukolsky operator for generically spinning BHs with beyond-Einstein couplings remains an open problem, this modified wave generation is the only piece we truncate. As detailed in the Supplementary Material, we can robustly bound the error introduced by this truncation. The omitted term scales as a multiplicative correction to the radiative rates ($\sim 5.2\%$ for dCS and $18\%$ for sGB for $r_0=3M$). Crucially, this correction is dynamically degenerate with the mass ratio $\eta$. Therefore, its omission does not fundamentally alter the physical phasing behavior, nor does it severely bias the recovery of the coupling parameter $\zeta$ in our Fisher analysis.

\sect{Effective Gravitational-Wave Signals from Scalar-Hair Modified Transition-to-Plunge}We model the \textit{effective} GW signals from a scalar-charge-modified orbit. This amounts to solving the Teukolsky equation~\cite{Teukolsky1972, Teukolsky1973, Teukolsky1974, Hartle_1974, Chandrasekhar_1975, MST_1996, MST_1997, Drasco_2004, Drasco_2006} along a shifted worldline due to the scalar charge
\begin{equation}
\hat{\mathcal{D}} [\psi] = \hat{S}[\mathcal{T}_{\alpha\beta}],\quad \mathcal{T}_{\alpha\beta} = \frac{m_s}{\sqrt{-g}} u_{\alpha} u_{\beta} \frac{\dd\tau}{\dd t} \delta^3(\mathbf{x} - \mathbf{x}_p(t)).
\end{equation}
where $\hat{\mathcal{D}}$ represents the full Teukolsky operator, $\psi=(r-ia\cos\theta)^4\psi_4$ is the rescaled Weyl scalar, and $\hat{S}$ is the source term encoding the EMRI dynamics under the combined influence of the 0PA-1SF self-force and the scalar radiation. In the definition of the point-particle energy-momentum tensor $\mathcal{T}_{\alpha\beta}$, $m_s$ denotes the mass of the secondary, $g$ is the determinant of the background metric, $u_{\alpha}$ is the particle's 4-velocity, $\tau$ is the proper time, $t$ is the coordinate time, and $\delta^3(\cdot)$ is the 3-dimensional Dirac delta function evaluating at the particle's trajectory along the scalar-charge shifted worldline $\mathbf{x}_p(t)$.

Strictly speaking, to model the waveforms consistently up to first order in the coupling parameter $\zeta$, one must solve this modified Teukolsky equation~\cite{Yang2026, Li2023}. However, to the best of our knowledge, the exact form of $\hat{\mathcal{D}}$ for generically spinning BHs with beyond-Einstein couplings remains an open problem. 
Therefore, in this work, we approximate the wave operator using its GR counterpart, $\hat{\mathcal{D}}^{(0)}$, solved in the frequency domain via the SN formulation~\cite{SN_1982, SN_1982b, Hughes_2000, Lo_2024, Lo2026, Yin2026}. 
While the resulting waveforms are not strictly accurate to linear order in $\zeta$, they successfully capture the leading-order waveform morphology. This approach provides valuable insight into how scalar charges of various coupling types and strengths affect the overall dynamics of the transition-to-plunge and merger phases.
With this approximation, the scalar-charge effects enter through the orbital phase, $\phi(t)$ being that of the deformed-potential worldline; the mode amplitudes are the corresponding Kerr ones. 
\begin{figure}[t]
\includegraphics[width=\columnwidth]{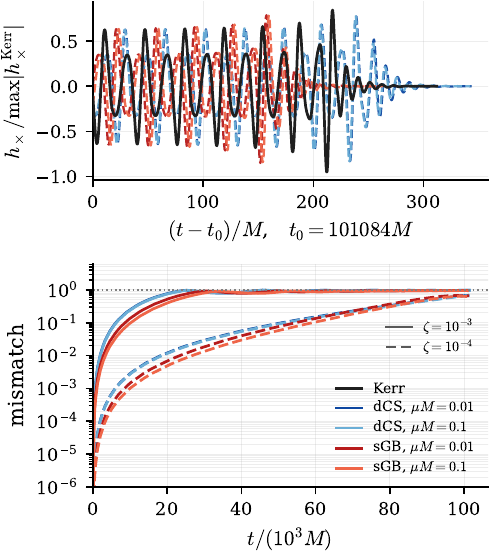}
\caption{\label{fig:dephasing}\emph{Top}: Gravitational waves of the cross-mode polarization $h_{\times}$ near merger with non-minimal coupling parameter of the massive scalar charge $\zeta=10^{-4}$, normalized to the Kerr peak, with relativistic phase and the circular-equatorial Teukolsky $(2,2)$-, $(3,3)$-, and $(4,4)$-amplitude continued by a $(2,2,0)$ ringdown (c.f. Supplementary Material). \emph{Bottom}: Mismatch between the Kerr transition-to-plunge waveform accumulated over $[0,t]$. The trajectories start at $r_{0}=5M$ at $a=0.8M$, $\eta=10^{-4}$ and are followed through $\approx\!1.7\times10^{3}$ orbits, viewed at an inclination $\iota=\pi/4$.}
\end{figure}

Equatorial orbits radiate only into modes with $\ell+m$ even, and we retain $(2,2)$, $(3,3)$ and $(4,4)$, normalizing the summed flux to the Kerr value so that the strain amplitude, and hence the signal-to-noise ratio, is unbiased. The higher multipoles are not a refinement here, as the $(2,2)$-mode supplies $78\%$ of the flux at $r=4M$, but only $30\%$ near the light ring (c.f. Supplementary Materials), and modes oscillating at different $m\Omega$ cannot all be realigned by a single time shift, which is what makes the imprint survive phase maximisation. Near the ISCO, the plunge is a transient, which means the spectrum is continuous and the frequency-domain Teukolsky amplitude at infinity, $Z^{\infty}_{\ell m}(\omega)$, follows from a convolution over the worldline. In the inspiral, the orbital evolution is adiabatic, and the amplitude is needed only at the harmonic $\omega=m\Omega(r)$, which we tabulate once and interpolate in the orbital frequency $\Omega$. Details on these calculations are summarized in the Supplementary Materials.

We compare the effective waveforms for EMRIs with an initial orbital separation $r_{0}=5M$. As seen in FIG.~\ref{fig:dephasing}, the effective correction accumulates over the whole radiation-reaction timescale, giving $\delta\varphi\simeq+21.3$ and $-45.2$~rad at $\zeta=10^{-3}$, falling to $+2.14$ and $-4.53$~rad at $\zeta=10^{-4}$ and scaling linearly in $\zeta$ between them. The mismatch against a Kerr template, maximised over time and phase and weighted by the LISA noise curve, is $0.386$ (dCS case) and $0.365$ (sGB case) at $\zeta=10^{-3}$, and $0.181$ and $0.114$ at $\zeta=10^{-4}$, all far above the nominal threshold $\sim D/2\rho^{2}\simeq5\times10^{-4}$, with $D$ the number of intrinsic waveform parameters~\cite{Lindblom_2008}, at ${\rm SNR}\simeq80$. This large mismatch is due to the inclusion of higher multipoles, as a single time shift absorbs most of the dephasing when only considering the $(2,2)$ mode. In this case, the $\zeta=10^{-4}$ mismatch collapses to $\sim10^{-3}$. The signature is therefore carried by when the secondary reaches the light ring. Specifically, the trajectory influenced by sGB coupling plunges first, dCS plunges last, Kerr in between.

\sect{Constraints on the Scalar Mass and Charge}We estimate how well LISA could measure $(\zeta,\mu M)$ from a single scalar-charge-modified transition-to-plunge signal, using a Fisher matrix analysis. 
The waveforms presented above carry the scalar radiation channel but not the $\mathcal{O}(\zeta)$ correction to wave generation. We take $M=10^{6}M_\odot$, $\eta=10^{-4}$, $a=0.8M$, a source at $d_L=1$~Gpc, and the SciRDv1 LISA sensitivity including the galactic confusion foreground~\cite{Robson2019}, as these are typical EMRI parameters~\cite{Babak_2007, Babak2017}. The $t\sim\!10^{5}M$ evolution from $r_{0}=5M$ lasts $\simeq6$~days and carries ${\rm SNR}\simeq80$. 
The derivative $\partial h/\partial\mu$ is evaluated by central differences from a dedicated pair of background solutions at $\mu M=0.200\pm10^{-3}$, and $\partial h/\partial\zeta$ from a $5\%$ variation about each fiducial. The Fisher-matrix analysis is performed over the parameters $(\zeta,\mu M,\ln M,\ln\eta,\ln A,t_{c},\phi_{c})$, and every quoted error is marginalised over the last five. That is, each quoted error is the square root of the corresponding diagonal element of the inverted $7\times7$ Fisher matrix, so that uncertainty in the remaining five parameters, and their correlations with $(\zeta,\mu M)$, propagate into the scalar errors rather than being assumed away. The forecast is therefore driven neither by arrival time alone nor by assumed knowledge of the binary parameters. The spin is held fixed, as it may be determined independently of a scalar-charge search by X-ray reflection/reverberation~\cite{Reynolds97, Reynolds99, Reynolds13, Reynolds21, Dauser10a, Dauser10b, Dauser13, Bambi21, Lui2026b} or GWs from the early inspiral~\cite{Burke2020, Chapman_Bird_2023, Vazquez_Aceves_2025}. Furthermore, it has been demonstrated that the spin can be measured even when submersed in a cloud of ultralight bosons~\cite{Chung2021}.
\begin{figure}[t]
\includegraphics[width=\columnwidth]{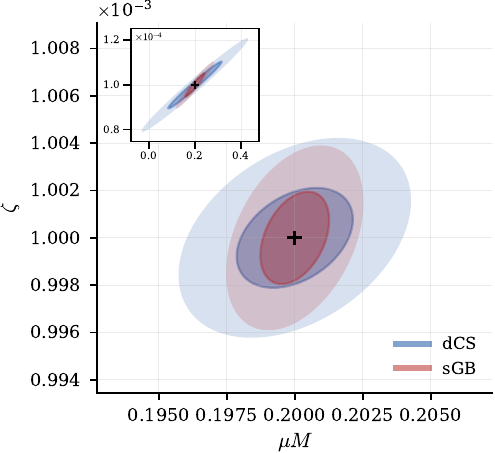}
\caption{\label{fig:fisher}Marginalised $1\sigma$ and $2\sigma$ Fisher contours on $(\mu M,\zeta)$ for a $10^{6}M_\odot$--$10^{2}M_\odot$ EMRI at $a=0.8M$ and $d_{L}=1$~Gpc (${\rm SNR}\simeq80$), using the three modes $(2,2)+(3,3)+(4,4)$. The main panel is $\zeta_{\rm fid}=10^{-3}$, the inset $\zeta_{\rm fid}=10^{-4}$. Errors are marginalised over $\ln M$, $\ln\eta$, $\ln A$, $t_{c}$ and $\phi_{c}$.}
\end{figure}

Figure.~\ref{fig:fisher} shows the $1\sigma$ and $2\sigma$ contours, forecasted with GWs of the $(2,2)$-, $(3,3)$- and $(4,4)$-modes. At $\zeta=10^{-3}$ the coupling is measured to $\sigma_{\zeta}/\zeta \simeq 2.1\times10^{-3}$ for dCS coupling and $2.0\times10^{-3}$ for sGB coupling, and the scalar mass to $\sigma_{\mu}/\mu\simeq 1.1\times10^{-2}$ and $6.3\times10^{-3}$, respectively. We find that adding the higher multipoles improves $\sigma_{\zeta}$ by only a factor $1.4$ (dCS) to $1.5$ (sGB), but improves $\sigma_{\mu}$ by factors of $11.3$ (dCS) and $7.8$ (sGB), as modes with different $\ell$ gauge the deformation strength at different radii. It is this that breaks the $\zeta$--$\mu$ degeneracy. Treating $M$ and $\eta$ as unknown rather than fixed inflates the scalar-parameter errors by a factor $1.2$--$2.4$, the larger values applying to $\sigma_{\mu}$. At $\zeta=10^{-4}$ the leverage is lost, $\sigma_{\zeta}/\zeta$ degrades to $0.053$--$0.11$ and $\mu$ becomes unconstrained, so measuring the scalar mass requires $\zeta\gtrsim10^{-3}$ at an SNR of $\simeq80$.

\sect{Discussion and Conclusion}We modelled the scalar-hair transition-to-plunge self-consistently, building the circular sequence, ISCO, transition coefficients, and plunge from the exact radial potential of the deformed metric. The resulting, gauge-invariant shift of the ISCO frequency distinguishes the two couplings by sign as well as magnitude, and is a physical probe of the scalar charge.

The same shift accumulates over the radiation-reaction timescale into an $\mathcal{O}(10)$~rad GW dephasing, with the transition adding an endpoint signature rather than being where the effect switches on. Resolving multiple GW modes is necessary rather than merely advantageous, as the modes oscillating at different $m\Omega$ cannot all be realigned by a single shift in arrival time. Multimode GWs increase the phase-and-time-maximised mismatch by two orders of magnitude at $\zeta=10^{-4}$ and tighten $\sigma_{\mu}$ by an order of magnitude, because modes at different $m\Omega$ cannot be simultaneously absorbed by a time shift. 
Our waveforms carry the scalar radiation channel but not the modified-Teukolsky corrections. We conjecture that including these corrections would not drastically alter our results and, if anything, could tighten the constraints on the scalar charge, making our forecast conservative.

Our work opens up a new direction of research. The most immediate follow-up work is to extend the analyses to non-equatorial orbits. To further increase the accuracy of our probe, we should also consider the influence of the gravity of the massive scalar charges on the GWs emitted by the inspiral and the plunge. This will require significant theoretical development to formulate and solve the modified Teukolsky equation~\cite{Yang2026, Li2023} for generically spinning BHs and computing the quasinormal modes of the host BH surrounded by massive scalar charges, the latter of which is in progress using our quasinormal-mode code, \textsc{METRICS}~\cite{Chung2024, Chung2025}. One can also extend our studies to configurations of massive scalar charges without axisymmetry, which can be formed by superradiance of massive scalar fields with non-minimal curvature couplings~\cite{Alexander:2022avt}. While other astrophysical observations, such as ground-based GW detections and BH spin measurements, have already begun to constrain the parameter space of massive scalar charges~\cite{Barack:2018yly, Berti:2015itd, Yunes:2024lzm, Baryakhtar:2020gao}, our work demonstrates that future EMRI detections will provide a highly sensitive, complementary tool, as our model shows that isolating the conservative backreaction on the EMRI trajectory can provide stringent constraints on the scalar field. EMRIs also reach a different corner of the parameter space, as $\mu\hbar\simeq10^{-18}$--$10^{-17}$~eV for $\mu M=0.01$--$0.2$ at $M=10^{6}M_{\odot}$, sit in an ultralight band where the hair is sourced directly by curvature, so the spin-down bounds derived from superradiance do not apply to it.

\sect{Acknowledgments}L.L. would like to thank Nicolas Yunes and Dongjun Li for the insightful discussions. L.L. and A.T.O. were supported by the Beijing Natural Science Foundation (No. IS25014). A.K.W.C.~acknowledges the Herchel Smith Fellowship at the University of Cambridge for support of this work.

\bibliography{main}
\onecolumngrid
\appendix*
\section{Supplementary Materials}
\twocolumngrid

\setcounter{equation}{0}
\subsection{Metric modified by scalar charge}
The components of the background Kerr metric in modified Boyer-Lindquist coordinate $(t,r,\chi,\phi)=(t,r,\cos\theta,\phi)$ are
\begin{equation}\label{eq:kerr_bg}
\begin{aligned}
g_{tt}^{(0,0)} &= -\left(1-\frac{2Mr}{\Sigma}\right),\\
g_{t\phi}^{(0,0)} &= -\frac{2M^{2}ar}{\Sigma}(1-\chi^{2}),\\
g_{rr}^{(0,0)} &= \frac{\Sigma}{\Delta}, \qquad
g_{\chi\chi}^{(0,0)} \;=\; \frac{\Sigma}{1-\chi^{2}},\\
g_{\phi\phi}^{(0,0)} &=(1-\chi^{2})\left[r^{2}+M^{2}a^{2}\right.\\
&\left.\qquad+\frac{2M^{3}a^{2}r}{\Sigma}(1-\chi^{2})\right],
\end{aligned}
\end{equation}
where $\Sigma = r^2 + M^2 a^2 \chi^2$, $\Delta = (r-r_+)(r-r_-)$, with $r_{\pm}$ being the coordinate radius of the outer and inner event horizon. The perturbed metric can be written as a 2-parameter expansion in the mass ratio $\eta$ and the dCS/sGB coupling parameter $\zeta$ as shown in Eq.~\eqref{eq:metric_PS}. Following the prescription of Ref.~\cite{Chung2026}, we assume an Ansatz of the metric modified by scalar charge, $g^{(0,1)}_{\alpha \beta}$, can be written in the following form
\begin{equation}\label{eq:metric}
\begin{split}
g_{\alpha\beta}^{(0,1)}\dd x^{\alpha}\dd x^{\beta}
={}&-H_1\,\dd t^{2}+2H_2\,g_{t\phi}^{(0,0)}\,\dd t\,\dd\phi\\
&+H_3\left(g_{rr}^{(0,0)}\dd r^{2}+g_{\chi\chi}^{(0,0)}\dd\chi^{2}\right)\\
&+H_4\,g_{\phi\phi}^{(0,0)}\,\dd\phi^{2}.
\end{split}
\end{equation}
where the functions $H_i(r,\chi)$ encode the entire deformation with $H_1$ the redshift, $H_2$ the frame dragging, $H_3$ the $r$--$\chi$ block and $H_4$ the axial circumference. They are stationary ($\partial_t H_i=0$) and axisymmetric ($\partial_\phi H_i=0$) by construction, which is the property responsible for the conservative structure derived below. They are obtained by solving the coupled scalar and modified-Einstein equations with the spectral methods of Refs.~\cite{Lam2025, Lam2026, Chung2024, Chung2025, Chung2026}, once for each theory, and enter the equations discussed in this work through the connection coefficients $\Gamma^{\alpha\,(0,1)}_{\alpha\beta}$ built from $g^{(0,1)}_{\mu\nu}$ and its first derivatives. We evaluate the contravariant forcing $f^{\alpha}_{(0,1)}=-\Gamma^{\alpha\,(0,1)}_{\rho\sigma}u^{\rho}_{(0,0)}u^{\sigma}_{(0,0)}$ directly from tabulated $H_i$ for the dCS and sGB solutions.

Two features of Eq.~\eqref{eq:metric} matter for what follows. First, the deformation preserves both Killing vectors $\partial_t$ and $\partial_\phi$, so the spacetime remains stationary and axisymmetric at $\mathcal{O}(\zeta)$. Second, the $H_i$ are even in $\chi$ --- in dCS the pseudoscalar itself is odd, but the metric deformation is quadratic in it --- so the deformed spacetime retains the reflection symmetry $\chi\to-\chi$ and the equatorial plane remains a totally geodesic surface.

\subsection{Orbits of the deformed spacetime}

The deformation $g^{(0,1)}_{\mu\nu}$ is stationary and axisymmetric, so both
Killing vectors of the background survive it~\cite{Destounis_2021, Destounis2021b}.
The orbital energy and angular momentum,
\begin{equation}\label{eq:fluxes}
\E = -g_{t\alpha}u^{\alpha}, \qquad
\AM = g_{\phi\alpha}u^{\alpha},
\end{equation}
with the indices lowered by the \emph{full} metric of Eq.~\eqref{eq:metric}, are
therefore exact integrals of the deformed spacetime, not merely constants to
$\mathcal{O}(\zeta)$. This is what allows us to avoid a perturbative treatment
of the worldline altogether: rather than expanding $u^{\alpha}$ order by order
and integrating a forced geodesic on the Kerr background, we integrate
\emph{exact} geodesics of the full metric, with $\E$ and $\AM$ its own conserved
quantities.

Concretely, the equatorial radial potential is obtained from the exact
normalisation $u^{\alpha}u_{\alpha}=-1$ of the deformed metric,
\begin{equation}\label{eq:Rdef}
V_r(r) = r^{4}\,(u^{r})^{2}, \qquad
\Big(\frac{\dd r}{\dd\lambda}\Big)^{2} = V_r(r),
\end{equation}
in Mino time, and the whole worldline is built from it: circular orbits from
$V_r=\partial_r V_r=0$, the innermost stable circular orbit from
$\partial_r^{2}V_r=0$, and the plunge by integrating $\dd r/\dd\lambda=-\sqrt{V_r}$.
These are set out in the two subsections below. At $\zeta=0$, Eq.~\eqref{eq:Rdef}
reduces identically to the usual Kerr potential.

Working with the exact potential is not merely tidier. Linearising instead --
expanding the Kerr potential about the deformed ISCO -- leaves $\partial_r R$ and
$\partial_r^{2}R$ non-zero at $\mathcal{O}(\zeta)$ there, and so discards terms of
the same order as the effect being measured, precisely where the transition
expansion is most delicate.

\subsection*{The \texorpdfstring{$\mathcal{O}(\zeta)$}{O(zeta)} force and why it is conservative}

It is still useful to exhibit the deformation as a force on the background, both
to see its size along the worldline and to establish that it does no secular
work. Splitting only the connection of the full metric and moving the
perturbation to the right-hand side gives, with no approximation,
\begin{equation}\label{eq:forceexact}
u^{\nu}\nabla^{(0)}_{\nu}u^{\alpha}
= -\,\Gamma^{\alpha\,(0,1)}_{\rho\sigma}\,u^{\rho}u^{\sigma}
\equiv f^{\alpha}_{(0,1)},
\end{equation}
where $\nabla^{(0)}$ is the background connection and $u^{\alpha}$ is the full
4-velocity. Because the 4-velocity is not expanded, the rearrangement is
closed: no cross terms are generated and the entire effect of the deformation is
the single term on the right. We evaluate $f^{\alpha}_{(0,1)}$ by
differentiating the exact connection of the deformed metric with respect to
$\zeta$ at $\zeta=0$ and contracting it with the full $u^{\alpha}$; this is the
quantity plotted in the lower panel of FIG.~\ref{fig:radiation}.

\begin{figure}[t]
\includegraphics[width=\columnwidth]{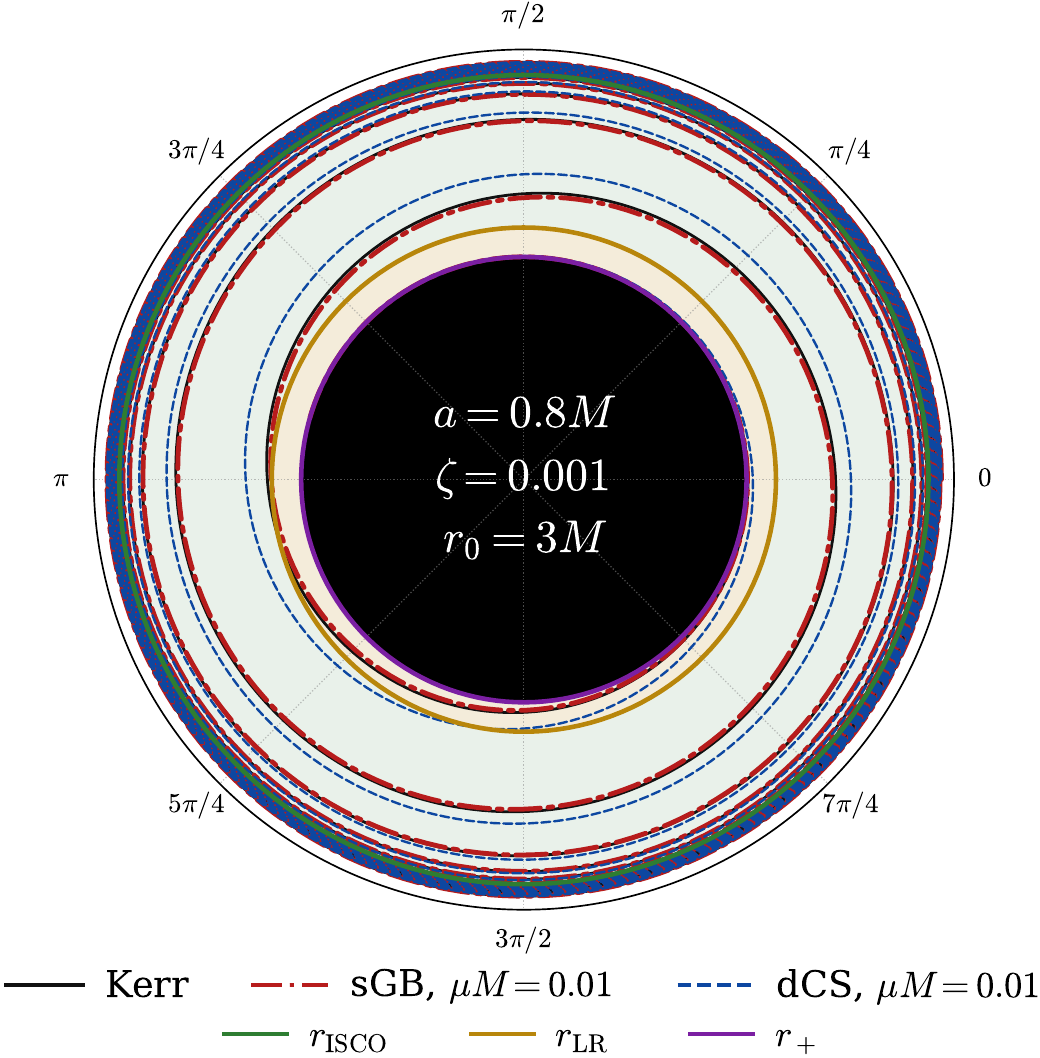}
\caption{\label{fig:inspiral+waveform}Equatorial transition-to-plunge trajectories into a rotating BH ($a=0.8M$), released from a common initial radius $r_{0}=3M$. The plot overlays three cases to compare the standard Kerr spacetime (solid black) with sGB (dash-dotted blue) and dCS (dashed red) models. Both modified gravity cases feature a coupling parameter $\zeta=10^{-3}$ and a scalar mass $\mu M=0.01$. The green, gold, and purple rings denote the innermost stable circular orbit (ISCO), light ring, and event horizon for the Kerr BH, respectively.}
\end{figure}

Figure.~\ref{fig:inspiral+waveform} summarizes the EMRI system we study in the main text. The scalar-charge-induced force is conservative for circular orbits, so it is dynamically suppressed through the adiabatic inspiral and starts to dominate once the non-adiabatic radial motion begins at the plunge. The scalar-charge imprint is therefore concentrated in last $\mathcal{O}(100)$ cycles before merger, where the trajectory is least contaminated by environmental effects, such as drag/torque from gas accretion~\cite{Duque_2025, Duque2025b, Duque2026, Lui2025, Lui2026, Hegade2025, Hegade2025b, Dyson_2026, Kejriwal2024, Kejriwal2026} other dark matter environments~\cite{Dai_2022, Dai_2024, Mitra2025, Dyson_2025, Li2025, Rahman2026}, and tidal resonance from other EMRI secondaries~\cite{Chen2018, Yang2017, Yang2019, Bonga2019, Cardoso2021, Silva2022, Silva2025, Yin2024, Santos2025, Santos2026}.

Because $\E$ and $\AM$ are exact constants of the deformed metric, the force does
no secular work: for circular equatorial orbits
$f^{t}_{(0,1)}=f^{\phi}_{(0,1)}=0$ pointwise, and once $u^{r}\neq0$ the individual
components are non-zero but represent the mismatch between the Kerr and deformed
constants rather than a flux. The deformation is therefore purely conservative
at $\mathcal{O}(\zeta)$, and $\E$ and $\AM$ evolve only through radiation
reaction.

\subsection*{Dissipative evolution of the constants}

The constants drift under the radiated fluxes, which enter at
$\mathcal{O}(\eta)$ and $\mathcal{O}(\eta\zeta)$:
\begin{equation}\label{eq:drift}
\frac{\dd\consts}{\dd t} = -\eta\,\dot{\consts}\big(r_{\rm eff}\big)\big[1+c(r)\,\zeta\big], \qquad \consts\in\{\E,\AM\}.
\end{equation}
Here $\dot{\consts}$ is the corresponding Kerr circular flux, since $r$ is gauge-dependent while the orbital phase is invariant, making the asymptotic frequency $\Omega$ a robust observable. This displacement of the evaluation point is
the part of the $\mathcal{O}(\eta\zeta)$ cross term that we retain exactly. The
coefficient $c(r)$ carries the remainder that we compute: the chain-rule response
of the flux to the deformation, together with the scalar radiation channel of the
following section. The piece we do not compute is the $\mathcal{O}(\zeta)$
correction to wave \emph{generation}, bounded below.

The adiabatic inspiral then follows from Eq.~\eqref{eq:drift} as
\begin{equation}
\frac{\dd r}{\dd t} = \frac{\eta\,\dot{\E}(r_{\rm eff})\,[1+c(r)\zeta]}{\dd\E/\dd r},
\qquad
\frac{\dd\varphi}{\dd t} = \Omega(r),
\end{equation}
with $\E(r)$ and $\Omega(r)$ taken from the exact circular sequence of the
deformed metric.

\subsection*{Remarks on the dissipationless at \texorpdfstring{$\mathcal{O}(\zeta)$}{O(zeta)} and bounds on the truncated cross term}The forcing $f^{\alpha}_{(0,1)}$ is built entirely from $g^{(0,1)}_{\mu\nu}$, which is stationary and axisymmetric. Both Killing vectors of the background therefore survive the deformation~\cite{Destounis_2021, Destounis2021b}, so $\E$ and $\AM$ as defined in Eq.~\eqref{eq:fluxes} remain exact constants of the \emph{deformed} metric. For circular equatorial orbits this is realised pointwise, $f^{t}_{(0,1)}=f^{\phi}_{(0,1)}=0$; once $u^{r}\neq0$ the individual components are non-zero but represent the mismatch between the Kerr and deformed constants rather than a secular flux, and average to zero over a radial period for bound motion.

Consequently, the primary's charge can influence dissipation only by modifying radiation sourced by the secondary. That radiation is $\mathcal{O}(\eta)$, so each of these channels, the GW flux computed on the deformed background via the modified Teukolsky equation~\cite{Li2025, Li2026}, horizon absorption by the hairy primary, and scalar radiation excited as the secondary moves through the primary's field, is $\mathcal{O}(\eta\zeta)$. There is no $\mathcal{O}(\zeta\eta^{0})$ dissipative channel available.

All such terms enter the transition equation in Eq.~\eqref{eq:transition} through a single scalar, the radiative coefficient $B$. Writing the flux on the deformed background as
\begin{equation}\label{eq:fluxsplit}
\dot{\consts}[\consts;\zeta]=\underbrace{\dot{\consts}_{\rm GW}(\consts+\delta\consts)}_{\text{chain rule}}
+\underbrace{\zeta\left(\frac{\partial \dot{\consts}}{\partial\zeta}\right)_{\consts}}_{\text{modified Teukolsky}},
\end{equation}
the two pieces differ enormously in cost. The first is fixed by $\delta\dot{\consts}=(\partial\dot{\consts}/\partial\consts)\,\delta\consts$: the hair displaces the orbit, and one evaluates the \emph{known} Kerr flux function at the displaced orbit. No new wave equation is involved, so we compute it exactly and retain it. The second requires the metric perturbation on the deformed background --- the modified Teukolsky problem, in which separability is lost and an effective source built from $g^{(0,1)}_{\mu\nu}$ must be reconstructed --- and is truncated.

The chain-rule piece is not a small correction. At $a=0.8M$, $\zeta=10^{-3}$, $\mu M=0.01$, the ISCO moves \emph{inward} by $\delta r/r=-5.63\times10^{-4}$ (dCS), while the Kerr circular flux near the ISCO is steep, $\dd\ln\dot{\E}/\dd\ln r=-4.70$ (stable to $2\%$ over $2.91\le r/M\le3.21$). The flux therefore shifts by $+0.26\%$, an effective $c\simeq+2.6$, and the resulting change opposes the conservative shift. Since the Boyer--Lindquist coordinate $r$ is not gauge-invariant we evaluate $\kappa_{\consts}$ at the deformed ISCO's orbital frequency $\Omega$, solving $\Omega_{\rm Kerr}(r_{\rm eff})=\Omega_{\rm ISCO}^{(\zeta)}$; matching on $r$ instead shifts $\Delta\phi$ by only $0.6\%$, which measures the residual gauge ambiguity.

Two consequences deserve emphasis. First, the comparison must be made between worldlines released from a \emph{common} radius. Releasing each theory at its own ISCO subtracts the dominant physical effect --- that the ISCO has moved --- leaving a small residual that the chain-rule term can overwhelm and whose sign it can reverse. With a common start, the signal is an order of magnitude larger and the corrections are shown in TABLE.~\ref{tab:truncation}.
\begin{table}[htbp]
\caption{\label{tab:truncation}Transition-to-plunge shifts for worldlines released
from a common radius $r_{0}=3M$, at $\eta=10^{-4}$, $\mu M=0.01$, $\zeta=10^{-3}$. The last column is the raw size of the truncated term.}
\begin{ruledtabular}
\begin{tabular}{lccc}
 & $\Delta t_{\rm plunge}$ & $\Delta\phi$ [rad] & truncated $\mathcal{O}(\eta\zeta)$ \\
\hline
dCS & $+6.95M$ & $+1.117$ & $5.2\%$ \\
sGB & $-1.53M$ & $-0.220$ & $18\%$ \\
\end{tabular}
\end{ruledtabular}
\end{table}

In TABLE.~\ref{tab:truncation}, the worldlines share a common initial radius $r_0=3M$, $\eta=10^{-4}$, $\mu M=0.01$, and the last column is the explicit modified-Teukolsky piece, evaluated with the calibrated coefficients $c=1.32$ and $0.91$, respectively. Second, the truncated remainder is a few-percent effect for the parity-odd coupling but reaches $18\%$ for the parity-even one. This is not because the omitted term is larger there, but because the retained signal is smaller at this release radius, $\Delta\phi=-0.22$~rad. Quoted dephasings therefore remain sensitive to the release point, as in any inspiral comparison, and the reference condition must be stated alongside the number.

The truncated term is nonetheless benign, for a reason independent of its size. It enters only as a multiplicative correction to the radiative rates, $\delta\dot{\consts}/\dot{\consts}=c\,\zeta$,
and in both the adiabatic and the transition equations the mass ratio and those rates appear
only through the product $\eta\dot{\consts}$. The constant $c$ is therefore \emph{exactly} degenerate with $\eta\to\eta(1+c\zeta)$: we verify numerically that the two worldlines agree to $6\times10^{-9}$~rad in accumulated phase. Since the Fisher analysis marginalises over $\ln\eta$, the truncated piece cannot bias the recovery of $\zeta$; it can only degrade it, and that degradation is already contained in the quoted errors. Only the \emph{radial variation} of $c$ is non-degenerate. We calibrate $c$ without ever solving the modified Teukolsky equation. Within Kerr we measure how the circular flux responds to a fractional metric perturbation at fixed orbital frequency, using a change of spin to supply a perturbation of known size, and then multiply that response by the amplitude of our own deformation, $|\zeta H_{i}|$. This calibration gives a profile running from $c=0.55$ to $1.93$ over the radii of the $r_{0}=5M$ inspiral samples, $2.85\le r/M\le5.1$.
Evolving the same worldline with that varying $c(r)$, and again with a constant $c$ equal to
its mean, shifts the accumulated phase by $16.2$ and $13.9$~rad respectively, so all but
$2.3$~rad is reabsorbed by rescaling the mass ratio. The residual, $11\%$ of the
$21.3$~rad signal, degrades the quoted precisions but cannot bias them, since the Fisher
analysis marginalises over $\ln\eta$. The dephasing is thus robust; the forecast precisions are conditional on this residual.

Our work is different from the previous studies~\cite{Barsanti2026, Speri2026}, where the scalar charge belongs to the secondary and the imprint is dissipative, as their $-1$~pN flux is strongest at large separation, growing linearly with orbital radius in units of the primary mass and quadratically with the secondary's dimensionless scalar charge.
In our study, the primary acquires a scalar charge which deforms the background geometry rather than adding a source. By the argument above, this $\mathcal{O}(\zeta)$ term is purely conservative, shifting the ISCO and thereby advancing/retarding the orbital evolution up to the transit-to-plunge phase. 
The two mechanisms probe different bodies, and their effects dominate in different stages of the inspiral.

\subsection*{Circular equatorial orbits and the ISCO shift}

In the equatorial plane $\chi=0$ we have $\Sigma=r^{2}$, and a circular orbit has $u^{r}=u^{\theta}=0$, so $u^{\alpha}=(u^{t},0,0,u^{\phi})$. The $\theta$ equation is satisfied by the reflection symmetry noted above, and the $t$- and $\phi$-equations are satisfied because the deformation is stationary and axisymmetric. Only the radial equation carries content. Writing $g_{\mu\nu}$ for the equatorial components of the \emph{full} metric, the $r$-component of the forced geodesic equation together with the normalization condition gives the closed pair
\begin{equation}\label{eq:radbal}
\left\{
\begin{aligned}
 &-\tfrac12 g^{rr}\big[\partial_r g_{tt}\,(u^{t})^{2}
   +2\,\partial_r g_{t\phi}\,u^{t}u^{\phi}\\
 &\qquad\qquad
   +\partial_r g_{\phi\phi}\,(u^{\phi})^{2}\big] =f^{r}_{(0,1)},\\
 &\;\,g_{tt}(u^{t})^{2}+2g_{t\phi}u^{t}u^{\phi}
   +g_{\phi\phi}(u^{\phi})^{2} =-1,
\end{aligned}
\right.
\end{equation}
with $g^{rr}=\Delta/\Sigma=(r^{2}-2Mr+a^{2})/r^{2}$. Eq~\eqref{eq:radbal} is the statement that the hair-induced radial force is balanced by a shift in the orbital velocity. At $\zeta=0$, $f^r_{(0,1)}$ vanishes, and the Kerr circular orbit is recovered. We solve the pair of equations in Eq.~\eqref{eq:radbal} using Newton iteration, seeded with the Kerr values $\Omega_{\rm K}=\sqrt{M}/(r^{3/2}+a\sqrt{M})$, and read off
\begin{equation}
\E=-g^{(0,0)}_{t\alpha}u^{\alpha},\qquad
\AM=g^{(0,0)}_{\phi\alpha}u^{\alpha},
\end{equation}
and $\Omega=u^{\phi}/u^{t}$. The deformation is stationary and axisymmetric, so $\partial_{t}$ and $\partial_{\phi}$
remain Killing vectors of the \emph{full} metric~\cite{Destounis_2021, Destounis2021b}.
Contracting them with the 4-velocity Killing vector/tensors therefore gives quantities that are exactly conserved along geodesics of the deformed spacetime, not
merely to $\mathcal{O}(\zeta)$; this is the content of Eq.~\eqref{eq:fluxes}.

The ISCO is the marginally stable circular orbit, located by $\dd\E/\dd r=0$ and bracketed below by the prograde circular photon orbit, $r_{\rm ph}=2M\{1+\cos[\tfrac23\arccos(-a/M)]\}$~\cite{Bardeen_1972}, and above by $r=12M$. The lower bound is the standard Kerr result and is the smallest radius at which any circular orbit exists, as the upper bound is a safe bracket, since the ISCO never exceeds $9M$ (the
retrograde extremal value) for any spin. At $\zeta=0$ the solver reproduces the exact Kerr value $r_{\rm ISCO}=2.90664386M$ at $a=0.8M$ to $2.1\times10^{-9}$, which fixes the numerical error well below the deformation being measured. At $\zeta=10^{-3}$, $\mu M=0.01$ we obtain $r_{\rm ISCO}=2.90500821M$ (dCS) and $2.90664802M$ (sGB), i.e.\ $\delta r_{\rm ISCO}/r_{\rm ISCO}=-5.627\times10^{-4}$ and $+1.435\times10^{-6}$; at $\mu M=0.1$ they become $-5.659\times10^{-4}$ and $+3.960\times10^{-5}$, respectively.

The radial displacement is gauge dependent and nearly vanishes for the parity-even coupling; the corresponding shifts of the ISCO frequency are $\delta\Omega_{\rm ISCO}/\Omega_{\rm ISCO}=-2.331\times10^{-4}$ and $+6.113\times10^{-4}$ at $\mu M=0.01$, and $-2.129\times10^{-4}$ and $+4.471\times10^{-4}$ at $\mu M=0.1$. The frequency shifts have \emph{opposite sign} in the two theories. On the logarithmic axes
of the top panel of FIG.~\ref{fig:isco} the curves are power laws, $\delta r_{\rm ISCO}\propto
\zeta^{p}$, and for dynamical Chern-Simons and for scalar Gauss-Bonnet at $\mu M=0.1$ the
exponent stays at $p\simeq1$ across $10^{-6}\le\zeta\le10^{-2}$, i.e., strictly linear. The
exception is scalar Gauss-Bonnet at $\mu M=0.01$, whose linear coefficient is anomalously
small; there the $\mathcal{O}(\zeta^{2})$ term overtakes it above $\zeta\approx10^{-3}$ and
$p$ rises to $1.8$. The gauge-invariant frequency shift remains close to linear in every case
($p=1.00$--$1.05$), which is the statement the $\mathcal{O}(\zeta)$ truncation actually
requires.

\subsection*{Integrating the transition and the plunge}

The Kerr radial potential for equatorial motion is~\cite{Carter_1968, Fujita_2009}
\begin{equation}\label{eq:Rrad}
\begin{split}
V_r(r)= & (\E^{2}-1)r^{4}+2Mr^{3}+\big[a^{2}(\E^{2}-1)-\AM^{2}\big]r^{2}\\
& +2M(\AM-a\E)^{2}r,
\end{split}
\end{equation}
with $\dd r/\dd\lambda=-\sqrt{V_r}$ in Mino time. At the ISCO $V_{R}=\partial_r V_{R}=\partial_r^2 V_{R}=0$ simultaneously, so with $x=r-r_{\rm ISCO}$ the leading behaviour is cubic in $x$ and linear in the drift of the constants, giving Eq.~\eqref{eq:transition} with
\begin{equation}
\begin{split}
A& =-\tfrac14\left(\partial_r^{3}V_r\right)|_{r=r_{\rm ISCO}}, \\
B& =\tfrac12 \left.\left[\big(\partial^2_{r\E}\big)\kappa_{\E}+\big(\partial^2_{rL}V_r\big)\kappa_{\AM}\right]\right|_{r=r_{\rm ISCO}}
\end{split}
\end{equation}
Using $\partial_r^{3}R=24(\E^{2}-1)r+12M$ together with the ISCO conditions gives $\E^{2}-1=-2M/(3r_{\rm ISCO})$ and hence $A=M$ identically, for every spin --- a useful check on the ISCO solver, which we verify to an accuracy of $10^{-9}$. The transition is seeded in the matching region $x_0\ll r_{\rm ISCO}$ with the asymptotic adiabatic solution $\lambda_0=Ax_0^{2}/B$, $(dx/\dd\lambda)_0=B/(2Ax_0)$, and integrated with the constants drifting linearly, $\consts(\lambda)=\consts_{\rm ISCO}+\kappa_{\consts}\lambda$. Solving Eq.~\eqref{eq:transition} for a range of $\eta$ reproduces the expected width scaling $x\propto\eta^{2/5}$ of Refs.~\cite{Ori2000, Becker2025}.

The transition solution degrades once $x$ is no longer small, so we hand off to an exact plunging geodesic at $r_{\rm TP}=r_{+}+(1-\beta)(r_{\rm ISCO}-r_{+})$ with $\beta=0.35$, freezing the constants at their handoff values. It is essential to integrate the plunge using the \emph{first-order} constraint $\dd r/\dd\lambda=-\sqrt{V_r(r)}$ rather than the second-order radial equation: the latter is formally equivalent but does not enforce the mass-shell condition, and integrating it accumulates constraint violation that reaches $u^{\alpha}u_{\alpha}\simeq+64$ at the horizon instead of $-1$. With the first-order form, $u^{\alpha}u_{\alpha}=-1$ is preserved to machine precision all the way to $r_{+}+10^{-4}M$. Worldlines are always released from a common radius, but that radius differs by figure: $r_{0}=3M$ for FIGS.~\ref{fig:inspiral+waveform} and~\ref{fig:isco}, and $r_{0}=5M$ for FIG.~\ref{fig:dephasing} and the Fisher forecast, where a longer evolution is wanted. Each comparison is internally consistent; the release radius is stated with every quoted number.

\subsection*{Frequency-domain Teukolsky solution}

The worldline sources the $s=-2$ Teukolsky equation. Since the plunge is a transient rather than a periodic orbit, the emitted spectrum is continuous, and the mode amplitudes follow from a convolution of the source over the worldline,
\begin{equation}
\begin{split}
& Z^{\infty}_{\ell m}(\omega) =\int \;
\Big[A_{nn0}+A_{\bar m n0}+A_{\bar m n1}+A_{\bar m\bar m 0}+\dots\Big] \\
& \quad \quad \quad \times e^{i\omega t(\lambda)-im\phi(\lambda)}\;\dd\lambda,
\end{split}
\end{equation}
the $A$ coefficients being the standard Teukolsky source amplitudes built from $u^{\alpha}$ and the radial function. We solve the radial problem in the Sasaki-Nakamura (SN) form rather than the Teukolsky form directly, since the SN potential is short-ranged and admits stable numerical integration out to large radii. Three implementation points proved essential and are recorded here because each produces a large error when handled naively.

\emph{(i) Angular eigenvalue convention.} The SN potentials require the Teukolsky angular separation constant in the form $\lambda=A_{\ell m}-2ma\omega+(a\omega)^{2}$, where $A_{\ell m}$ is the spheroidal eigenvalue. Using the spin-weighted spherical combination $A+s(s+1)$ instead leaves the equations superficially well behaved but produces solutions that are wrong by tens of percent.

\emph{(ii) Radial variable.} The SN equation is naturally posed in the tortoise coordinate $r_{*}$, but its potentials depend on $r$. Recovering $r$ by interpolating a tabulated $r(r_{*})$ inside the right-hand side destroys the accuracy of the solution. We instead carry $r$ as an additional dependent variable, integrating $\dd r/dr_{*}=\Delta/(r^{2}+a^{2})$ alongside the SN system, which restores full solver accuracy.

\emph{(iii) Validation.} We check the homogeneous sector in absolute normalization against the independent \textsc{GeneralizedSasakiNakamura.jl} implementation~\cite{Hughes_2000, Lo_2024, Lo2026, Yin2026}, matching the ingoing solution $R^{\rm in}$ at $r=10M$ and the incidence amplitude $B^{\rm inc}$ to $\sim10^{-4}$. As an end-to-end test, the ringdown produced by the plunging worldline reproduces the Kerr $(2,2,0)$ quasinormal frequency at $a=0.8M$, $M\omega=0.5860-0.0756i$, to $0.3\%$ in the real part.

Our own source amplitudes retain a relative $\mathcal{O}(v^{2})$ error. Their leading order is correct --- the circular $(2,2)$ flux tends to the quadrupole result as $v\to0$ --- but the next term carries the wrong coefficient, $\dot E_{\rm Newt}(1+7.3v^{2})$ against the exact $(1-\tfrac{107}{21}v^{2})$ for $a=0$; summed over modes this makes the radiated power a factor $2.0$ too large near the ISCO and the strain $\sqrt{2}$ too large. Since the strain normalization sets the SNR, and Fisher errors scale as ${\rm SNR}^{-1}$, this alone would bias the forecast optimistically by $40\%$. We therefore do not use these amplitudes for the waveform: the phasing comes from the deformed-potential worldline, and the mode amplitudes are taken from the \textsc{FastEMRIWaveforms} Teukolsky tables, whose convention we fix by requiring $\sum_{\ell m}2\omega^{2}|Z_{\ell m}|^{2}$ to reproduce the corresponding Kerr flux; that ratio is constant to $0.4\%$ over $3.5\le p/M\le10$. Our SN solver is retained for the homogeneous sector and for the transient plunge of FIG.~\ref{fig:inspiral+waveform}. The \emph{differences} between $\zeta\neq0$ and $\zeta=0$ waveforms, which is what this work reports, are far better converged because the error is common mode and cancels in the comparison.

\subsection*{Waveform construction: transient and adiabatic regimes}

Two different constructions are needed, because the frequency-domain treatment above is designed for a \emph{transient}. Its cost is set by the frequency resolution: an inverse transform on a grid of spacing $\Delta\omega$ is periodic in time with period $2\pi/\Delta\omega$, so a signal of duration $T$ requires $\Delta\omega\ll2\pi/T$. For a source released near the ISCO, $T\sim\!5\times10^{2}M$ and a grid of a few hundred points suffices. For a source released at $r_{0}=5M$, $T\simeq10^{5}M$ and the same criterion would demand $\sim\!10^{4}$ frequencies per case. The worst part is that the long inspiral is quasi-monochromatic, so $Z^{\infty}_{\ell m}(\omega)$ approaches a distribution that no practical grid resolves.

\emph{Transient regime (FIG.~\ref{fig:inspiral+waveform}).} The worldline is released just outside the ISCO, at $r_{\rm ISCO}+0.06M$, and the full convolution is evaluated on a grid with $M\Delta\omega=5\times10^{-3}$, i.e.\ a transform period of $1.26\times10^{3}M$ against a signal of $\sim\!4\times10^{2}M$. Plunge and ringdown then emerge from the calculation itself, with no attachment. We stress that this margin is not optional: with $M\Delta\omega=10^{-2}$ the period drops to $6.3\times10^{2}M$, the ringdown wraps onto the inspiral, and the reconstructed strain develops a spurious burst that \emph{grows} after horizon crossing.

\emph{Adiabatic regime (FIG.~\ref{fig:dephasing}).} Over the long inspiral, the orbit is quasi-circular, so the Teukolsky amplitude is required only at the harmonic $\omega=m\Omega(r)$, which only requires one radial solve per orbit, not a continuum. We tabulate the circular-equatorial $Z^{\infty}_{\ell m}$ on a grid in $r$ and interpolate in the observable orbital frequency $\Omega$, so that a single table serves every scalar coupling and mass. The strain is then
\begin{equation}\label{eq:adstrain}
h_{+}-ih_{\times}= -\sum_{\ell m}{}_{-2}S_{\ell m}(\theta)\,
\frac{Z^{\infty}_{\ell m}\big(\Omega(t)\big)}{\big(m\Omega(t)\big)^{2}}\;e^{-im\phi(t)},
\end{equation}
with $\phi(t)$ being the orbital phase from the worldline. Since the orbital phase is gauge-invariant, making the asymptotic frequency $\Omega$ an unambiguous observable, the amplitude is evaluated consistently with the flux prescription used for $\kappa_{\mathcal{C}}$.

Only $\phi(t)$ carries the scalar hair, as the $Z^{\infty}_{\ell m}$ are Kerr amplitudes evaluated along the deformed worldline. This is consistent at the order we work, since their $\mathcal{O}(\zeta)$ correction is part of the truncated $\mathcal{O}(\eta\zeta)$ sector. Equatorial orbits radiate only into $\ell+m$ even modes, and we keep $(2,2)$, $(3,3)$, $(4,4)$; the summed flux is normalized to the Kerr value. The mode hierarchy collapses in the strong field --- the $(2,2)$ share of the flux is so that $(3,3)$ overtakes $(2,2)$ just outside the light ring. Since the modes oscillate at different $m\Omega$, no single time shift can realign them all. Consequently, the phase-and-time-maximised mismatch does not collapse.

Unstable circular orbits persist inside the ISCO all the way to the photon sphere, so the same table can be carried through the plunge. We extend it to $r=1.84M$, just outside $r_{\rm LR}=1.811M$, which requires supplying the analytic Kerr circular constants $\E(r)$, $\AM(r)$ directly. The table terminates at $M\omega=0.607$, within $4\%$ of the $(2,2,0)$ quasinormal frequency $M\omega=0.586$, showing the light-ring/quasinormal correspondence appearing here as a consistency check rather than an input. Beyond the light ring, the notion of a circular amplitude ends, and we continue the waveform with a Kerr $(2,2,0)$ ringdown matched in amplitude and phase at that point. This last step is an attachment, not a first-principles plunge; it is used only in FIG.~\ref{fig:dephasing}, whereas FIG.~\ref{fig:inspiral+waveform} carries a genuine Teukolsky merger.

Replacing the Teukolsky amplitude of Eq.~\eqref{eq:adstrain} by a Newtonian $(M\Omega)^{2/3}$ envelope changes the accumulated mismatches by $\lesssim0.2\%$, confirming that the amplitude is common mode between the $\zeta=0$ and $\zeta\neq0$ waveforms and that the dephasing results are phase driven.

\subsection*{The \texorpdfstring{$\mathcal{O}(\eta\zeta)$}{O(eta zeta)} scalar channel}

\emph{Construction.}
The secondary stirs the primary's hair, and the disturbed hair radiates. We
compute that channel following the treatment of an EMRI inside a scalar cloud
developed in Ref.~\cite{LiWeller2025}, adapted to hair that is sourced by a
curvature invariant rather than grown by superradiance. Both the scalar and the
metric are expanded in the mass ratio $\eta$ and the coupling $\zeta$,
\begin{align}
\Phi &= \zeta\Phi^{(1,0)} + \eta\,\zeta\Phi^{(1,1)} + \dots,\\
g_{\mu\nu} &= g^{(0,0)}_{\mu\nu} + \zeta^{2}h^{(2,0)}_{\mu\nu}
            + \eta\,h^{(0,1)}_{\mu\nu} + \dots,
\end{align}
where $\Phi^{(1,0)}$ is the stationary hair, $h^{(0,1)}_{\mu\nu}$ is the
gravitational perturbation sourced by the secondary, and $\Phi^{(1,1)}$ is the
scalar radiation we are after. One difference from Ref.~\cite{LiWeller2025} is
kinematic and simplifies matters: their cloud is a quasibound state with its own
frequency and azimuthal number $(\omega_{c},m_{c})$, whereas our hair is
stationary and axisymmetric, so $\omega_{c}=m_{c}=0$ everywhere.

\emph{The source term.}
Linearising the wave operator in $\eta$ gives
\begin{equation}\label{eq:scalsource}
\big(\Box^{(0,0)}-\mu^{2}\big)\Phi^{(1,1)}
 = -\,\delta\Box[h]\,\Phi^{(1,0)}\;-\;\alpha\,\delta\mathcal{I}[h],
\end{equation}
where the first term on the right-hand side is the linearisation of the wave operator itself,
\begin{align}
-\,\delta\Box[h]\,\Phi^{(1,0)}
 &= g^{\mu\nu(0,0)}\,\delta\Gamma^{\alpha}_{\mu\nu}\,\Phi^{(1,0)}_{,\alpha}
    \nonumber\\
 &\quad + h^{\mu\nu(0,1)}\,\nabla_{\mu}\nabla_{\nu}\Phi^{(1,0)}.
    \label{eq:dbox}
\end{align}
The two terms of Eq.~\eqref{eq:dbox} are exactly the source of
Ref.~\cite{LiWeller2025}. The remaining term $-\alpha\,\delta\mathcal{I}[h]$
has no counterpart there and is required here: their scalar is minimally
coupled, whereas ours is sourced by the curvature invariant itself,
$(\Box-\mu^{2})\Phi^{(1,0)}=-\alpha\mathcal{I}$, with $\mathcal{I}$ the
Pontryagin density for dCS and the Gauss-Bonnet density for
sGB. Perturbing the metric therefore perturbs the source that
creates the hair, at the same order $\eta\zeta$. Omitting it leaves a source
that is not gauge covariant. Two measurements fix the term with no free
parameters: the identity $\delta\mathcal{I}[\mathcal{L}_{\xi}g]
=\xi^{\mu}\partial_{\mu}\mathcal{I}$ is reproduced to an accuracy of $1.8\times10^{-5}$, and the
relative weight is $\alpha=1.0$ with a spread of $2\times10^{-5}$ across
$(r,\theta)$.

We evaluate both terms in the brace rather than dropping the first.
Reference~\cite{LiWeller2025} discards it because in Lorenz gauge
$g^{\mu\nu}\delta\Gamma^{\alpha}_{\mu\nu}=\nabla^{\mu}\bar h_{\mu\alpha}=0$. Our
reconstruction is also in Lorenz gauge, so the term is numerically negligible, but
retaining it is what allows the pure-gauge null test below to be run on modes
that do \emph{not} satisfy the Lorenz condition, where the cancellation cannot be
assumed.

\emph{Mode structure and fluxes.}
With $\omega_{c}=m_{c}=0$ the frequency selection rule of
Ref.~\cite{LiWeller2025} collapses to $\omega=m\Omega(r_{0})$. We solve the
radial equation by Green's function against the ingoing and outgoing homogeneous
solutions and evaluate
\begin{equation}
\dot E^{\,\infty}=\frac{\omega p}{4\pi}\big|Z^{\infty}\big|^{2},\qquad
\dot E^{\,H}=\frac{\omega k}{4\pi}\big(r_{+}^{2}+a^{2}\big)\big|Z^{H}\big|^{2},
\end{equation}
with $p=\sqrt{\omega^{2}-\mu^{2}}$ and $k=\omega-m\Omega_{H}$. These are the
fluxes of Ref.~\cite{LiWeller2025}, whose horizon normalization $4Mr_{+}$ equals
ours by $r_{+}^{2}+a^{2}=2Mr_{+}$. Modes with $\omega^{2}\le\mu^{2}$ are
evanescent and carry nothing to infinity, which is their threshold condition in
the $\omega_{c}=0$ limit; the horizon flux has no such threshold, since $k$ does
not involve $\mu$, and is \emph{negative} for $\omega<m\Omega_{H}$, where the
mode is superradiant, and the hole loses energy. We retain
$\ell\le9$ and verify that the $\ell$ sum has converged.

\emph{Parity.}
The selection rule inverts relative to Ref.~\cite{LiWeller2025}, whose source is
built from a scalar cloud. Ours is built from a curvature invariant, so the
radiating multipoles are set by the parity of that invariant: $\ell+m$ even for
the parity-even Gauss-Bonnet coupling, and the opposite half for the parity-odd
Pontryagin coupling, whose pseudoscalar reverses it. The non-radiating half is
suppressed by $9.7\times10^{-8}$ relative to the radiating one, which is a check
on the angular projection rather than an assumption.

\emph{The $m\ge2$ cut.}
The $m=1$ sector is excluded from both channels. Retaining it, $|c_{\rm hor}|$
\emph{grows} by an order of magnitude between $r_{0}=3M$ and $5M$
(TABLE~\ref{tab:mcut}), which is unphysical for a channel that must weaken as the
orbit widens; the sector is gauge dominated, in the sense that a pure-gauge
$h$ yields a nonzero converged flux there while the gauge-invariant 0th Weyl scalar $\psi_{0}$
falls with $r_{0}$ as it should. After the cut both $|c_{\rm inf}|$ and
$|c_{\rm hor}|$ fall monotonically at every mass and in both theories. The mass
threshold was already imposing this cut on $c_{\rm inf}$ implicitly, since
$\omega=\Omega<\mu$ for $m=1$ at the larger scalar masses.

\begin{table}[tbp]
\setlength{\tabcolsep}{3pt}\small\centering
\begin{tabular}{@{}lrrrrr@{}}
\hline\hline
$r_{0}/M$ & 3.0 & 3.5 & 4.0 & 4.5 & 5.0\\
\hline
\multicolumn{6}{@{}l}{\emph{sGB}}\\
$c_{\rm hor}$, all $m$ & $-2.446$ & $-4.341$ & $-8.838$ & $-19.22$ & $-37.73$\\
$c_{\rm hor}$ & $-0.876$ & $-0.538$ & $-0.350$ & $-0.232$ & $-0.160$\\
$c_{\rm inf}$ & $+2.133$ & $+0.695$ & $+0.275$ & $+0.126$ & $+0.065$\\
$c_{\rm tot}$ & $+1.257$ & $+0.157$ & $-0.075$ & $-0.106$ & $-0.095$\\
\hline
\multicolumn{6}{@{}l}{\emph{dynamical Chern-Simons}}\\
$c_{\rm hor}$, all $m$ & $-0.290$ & $-0.792$ & $-2.378$ & $-6.266$ & $-14.38$\\
$c_{\rm hor}$ & $-0.0871$ & $-0.0819$ & $-0.0687$ & $-0.0559$ & $-0.0453$\\
$c_{\rm inf}$ & $+0.0138$ & $+0.0032$ & $+0.0010$ & $+0.0003$ & $+0.0001$\\
$c_{\rm tot}$ & $-0.0733$ & $-0.0787$ & $-0.0678$ & $-0.0555$ & $-0.0451$\\
\hline\hline
\end{tabular}
\caption{\label{tab:mcut}Scalar-channel coefficients at $\mu M=0.01$, normalized
to the Kerr GW flux at the same $r_{0}$. Rows marked ``all
$m$'' retain every azimuthal mode; all others carry the $m\ge2$ cut.
$c_{\rm tot}=c_{\rm inf}+c_{\rm hor}$ is a sum of two monotone functions of opposite sign, so
$|c_{\rm tot}|$ does not need to be monotonic, and is not: for scalar
Gauss-Bonnet it passes through zero near $r_{0}\approx3.73M$, and for dynamical
Chern-Simons the near-cancellation is strongest at the innermost radius. The
interpolator uses a signed spline there rather than the $\log|c|$ power law used
elsewhere.}
\end{table}

\emph{One departure, and its size.}
We set the angular separation constant to its spherical value
$\Lambda_{\ell m}=\ell(\ell+1)$, whereas Ref.~\cite{LiWeller2025} uses the spin-0
spheroidal eigenvalue at spheroidicity $\gamma=a\sqrt{\omega^{2}-\mu^{2}}$ and
identifies the separation constant as one of two corrections it makes to earlier
work. We have measured the cost rather than estimated it. At the most demanding
point in our grid ($r_{0}=3M$, $m=2$, $\mu M=0.01$), the eigenvalue differs by at
most $2.2\times10^{-3}$ in relative terms, and the resulting fluxes by $1.5\%$ at
infinity and $0.35\%$ at the horizon; at $r_{0}=5M$ the eigenvalue shift falls to
$5\times10^{-4}$. Both are below the $4$--$6\%$ inner-cutoff convergence that
dominates our error budget, so we retain the spherical value and record the bias
here rather than absorbing it silently.

\begin{figure}[htbp]
\includegraphics[width=\columnwidth]{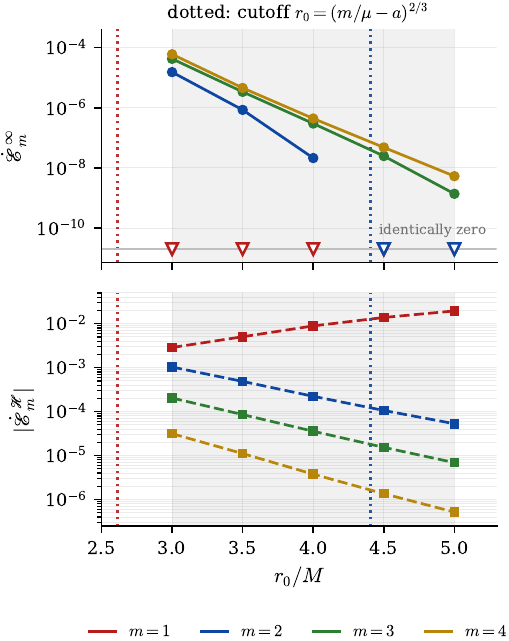}
\caption{\label{fig:livalid}Scalar flux in the $m=1$ to $m=4$ modes at
$\mu M=0.199$, $a=0.8M$, against the cutoff of Eq.~(53) of
Ref.~\cite{LiWeller2025} (dotted), which our mode bookkeeping must place
correctly in $r_{0}$. Colour labels $m$, for the curves and for their cutoffs
alike; only the $m=1$ and $m=2$ cutoffs fall in the plotted range, those for
$m=3$ and $m=4$ lying at $5.88M$ and $7.20M$, which is why those two modes
radiate at every radius shown. \emph{Top}: flux to future null infinity
$\mathscr{I}^{+}$. Open triangles on the floor mark modes that are identically
zero, which a logarithmic axis cannot otherwise show. \emph{Bottom}: flux down
the horizon $\mathscr{H}$ for the same modes, which carries no mass threshold
and is negative throughout, so its magnitude is plotted. The shaded band is the
range in which we solve.}
\end{figure}

\emph{A consistency check against Ref.~\cite{LiWeller2025}.}
Their Eq.~(53) fixes the orbital radius beyond which a given azimuthal mode
stops carrying energy to infinity. The mode turns evanescent once $|\omega|<\mu$,
which for stationary hair ($\omega_{c}=0$) becomes $r_{0}>(m/\mu-a)^{2/3}$. Their
horizon flux, Eq.~(61b), carries no such threshold, because the ingoing
wavenumber $k=\omega-m\Omega_{H}$ does not involve $\mu$. We adopt both rules, so
FIG.~\ref{fig:livalid} is a check of our mode bookkeeping rather than an
independent rederivation of them. What it tests is that the frequency map
$\omega=m\Omega(r_{0})$ places each cutoff at the radius their formula predicts,
and that the two channels switch off in different places as their framework
requires. At $\mu M=0.199$ the predicted cutoffs for $m=1$ and $m=2$ are $2.61M$ and
$4.41M$, one below our computed window $3M\le r_{0}\le5M$ and one inside it.
Our fluxes follow that pattern. The $m=1$ flux to infinity vanishes identically
at every radius, and $m=2$ carries flux at $r_{0}=3.0$, $3.5$ and $4.0M$ and
vanishes identically at $4.5$ and $5.0M$, straddling the predicted $4.41M$. The
cutoffs for $m=3$ and $m=4$ fall at $5.88M$ and $7.20M$, beyond the window, so
those modes radiate at every radius we solve and are not plotted. The horizon
flux, by contrast, is nonzero for both modes at all five radii and crosses each
cutoff without a feature, and is negative everywhere because absorption is
superradiant here. One caveat
is visible in the lower panel: the $m=1$ horizon flux \emph{grows} with $r_{0}$,
which is the unphysical behaviour that motivates the $m\ge2$ cut of
TABLE~\ref{tab:mcut}; it is shown here rather than hidden, and is excluded from
the channel we use.

\emph{Magnitude, and why the waveform is effective and conservative.}
The computed channel is small. Over $3M\le r_{0}\le5M$ at $\mu M=0.01$,
$|c_{\rm tot}|\le1.26$ for sGB and $|c_{\rm tot}|\le0.079$ for
dCS. It is therefore \emph{smaller} than the retained
chain-rule coefficient $c\simeq2.6$ --- by a factor of $2$ for the parity-even
coupling and up to $58$ for the parity-odd one. The dissipative sector is thus
dominated by a term we evaluate exactly, the Kerr flux at the displaced orbit,
rather than by the newly computed scalar radiation. The piece still missing, the
$\mathcal{O}(\zeta)$ modification of GW generation, would add
$\zeta$ dependence to the amplitudes that we do not model. Both considerations
point the same way: the waveforms here are \emph{effective}, and the dephasings,
mismatches and Fisher errors quoted in the main text should be read as an
\emph{effective} bound on the scalar charge.

One gap is stated explicitly. The sGB solution at $\mu M=0.1$
has no tabulated hair, so no flux table exists for it, and that single curve
carries the chain-rule coefficient alone. It is the only case in this work for
which the computed scalar channel is absent.

\subsection*{Normalisation of the scalar dipole flux}
For a secondary carrying squared dimensionless charge $\Lambda=d^{2}$, comparing the weak-field scalar dipole against the GW quadrupole gives~\cite{Barsanti2026, Speri2026}
\begin{equation}\label{eq:dipratio}
\frac{\dot E_{\rm s}}{\dot E_{\rm GW}}\simeq\frac{5}{384}\,\Lambda\,\hat r,
\end{equation}
with $\hat r=r/M$. This is a $-1$ pN effect, largest at wide separations and suppressed towards the ISCO, the two fluxes being comparable at $\hat r\simeq384/(5\Lambda)$. The coefficient depends on how the scalar kinetic term is normalized, and the factor is easy to lose. Reference~\cite{Barsanti2026} writes the action as $S_0=\int d^4x\sqrt{-g}\,(16\pi)^{-1}(R-\tfrac12\partial_a\varphi\partial^a\varphi)$, with the secondary entering through a scalar-dependent mass whose first-order coefficient is $m^{(1)}=-\mu d^{(0)}/4$. The field equation is then $\Box\varphi=-4\pi d\mu\int\delta^{4}(x-y_p)\,\dd\tau/\sqrt{-g}$, so the source strength is that of a charge $q=\mu d$ in the $\Box\varphi=-4\pi\rho$ convention. The radiated power, however, follows from the stress tensor of $S_0$, which carries the same $1/(16\pi)$. With $\oint\dot\varphi^{2}r^{2}\dd\Omega=(4\pi/3)|\ddot{\mathcal{D}}|^{2}$ for a dipole $\mathcal{D}=q\,x_p$, 
\begin{equation}
\dot E_{\rm s}=\frac{1}{16\pi}\cdot\frac{4\pi}{3}\,|\ddot{\mathcal{D}}|^{2}=\frac{|\ddot{\mathcal{D}}|^{2}}{12}
=\frac{\Lambda m_s^{2}M^{2}}{12\,r^{4}},
\end{equation}
using $|\ddot{\mathcal{D}}|^{2}=q^{2}\Omega^{4}r^{2}$ and $\Omega^{2}=M/r^{3}$. Dividing by $\dot E_{\rm GW}=(32/5)m_s^{2}M^{3}/r^{5}$ yields Eq.~\eqref{eq:dipratio}. A canonically normalized scalar would instead give $|\ddot{\mathcal{D}}|^{2}/3$ and hence $(5/96)\Lambda\hat r$, four times larger. We have checked the result against FIG.~1 of Ref.~\cite{Barsanti2026}: at $\Lambda=2.5\times10^{-3}$, Eq.~\eqref{eq:dipratio} gives $2.6\times10^{-4}$ at $\hat r=8$ and $9.8\times10^{-4}$ at $\hat r=30$, consistent with the plotted $\Lambda F_0^{\Lambda}/F_0^{\rm GSF}$ and with their statement that the scalar term becomes comparable to the first post-adiabatic gravitational term for $\hat r\gtrsim8$.

\subsection*{Coordinate behaviour inside the light ring}
Curves in the bottom panel of FIG.~\ref{fig:radiation} are truncated at the light ring. Inside it the Boyer--Lindquist components of $f^{\alpha}_{(0,1)}$, and the associated $\dd\consts/\dd\lambda$, grow as inverse powers of $\Delta$ and diverge at $r_{+}$. This is a property of the coordinate basis, not of the physics: $\Delta\to0$ at the horizon and Boyer--Lindquist time becomes singular there, exactly as it does for the unmodified Kerr geodesic. Scalar invariants built from $f^{\alpha}$ remain finite, and the worldline itself crosses the horizon smoothly with $u^{\alpha}u_{\alpha}=-1$ preserved. We therefore display the forcing only where its coordinate representation is meaningful, while integrating the trajectory through to $r_{+}$.

\end{document}